npj digital medicine

www.nature.com/npjdigitalmedREVIEW ARTICLE OPEN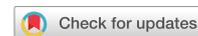

# Medication abortion via digital health in the United States: a systematic scoping review

Fekede Asefa Kumsa[1]✉, Rameshwari Prasad[1] and Arash Shaban-Nejad[1]✉Digital health, including telemedicine, has increased access to abortion care. The convenience, flexibility of appointment times, and ensured privacy to abortion users may make abortion services via telemedicine preferable. This scoping review systematically mapped studies conducted on abortion services via telemedicine, including their effectiveness and acceptability for abortion users and providers. All published papers included abortion services via telemedicine in the United States were considered. Articles were searched in PubMed, CINAHL, and Google Scholar databases in September 2022. The findings were synthesized narratively, and the PRISMA-ScR guidelines were used to report this study. Out of 757 retrieved articles, 33 articles were selected based on the inclusion criteria. These studies were published between 2011 and 2022, with 24 published in the last 3 years. The study found that telemedicine increased access to abortion care in the United States, especially for people in remote areas or those worried about stigma from in-person visits. The effectiveness of abortion services via telemedicine was comparable to in-clinic visits, with 6% or fewer abortions requiring surgical intervention. Both care providers and abortion seekers expressed positive perceptions of telemedicine-based abortion services. However, abortion users reported mixed emotions, with some preferring in-person visits. The most common reasons for choosing telemedicine included the distance to the abortion clinic, convenience, privacy, cost, flexibility of appointment times, and state laws imposing waiting periods or restrictive policies. Telemedicine offered a preferable option for abortion seekers and providers. The feasibility of accessing abortion services via telemedicine in low-resource settings needs further investigation.

npj Digital Medicine (2023)6:128 ; https://doi.org/10.1038/s41746-023-00871-2## INTRODUCTION

In recent years, the availability of intelligent digital health solutions has transformed the population and personalized healthcare[1]. The U.S. Food and Drug Administration (FDA) states that digital health is comprised of categories that include health information management (HIM) technology, mobile health (mHealth), personalized medicine, telehealth and telemedicine, and the use of wearable medical devices[2]. Telehealth and telemedicine allow the delivery of healthcare services such as counseling, assessment, and clinical guidance from a distance through electronic means of communication[3]. They also facilitate a personalized and targeted approach to improve patients' experience[4]. Regarding medication abortion services provided outside of inpatient healthcare settings before 12 weeks of gestational age, telemedicine and telehealth technologies present an alternative to in-clinic abortion services[5]. Previous studies have reported that the effectiveness of abortion services through telemedicine is comparable to that of in-person provision of abortion services[5–7]. The demand for medication and abortion services via telemedicine increased in the United States, particularly following the COVID-19 pandemic[8,9].

As the availability of telemedicine has increased, the ability of health providers to offer abortion services using this technology has varied widely. In some locations, only a limited number of abortion services, such as counseling or consenting, were available via telemedicine, while others were offered a wider range of services that included medication abortion[10–12]. The dispensing of the oral abortifacient mifepristone via telemedicine was originally permitted under an FDA investigational new drug application (INDA) protocol given the FDA restricted its provision via telemedicine and permitted its administration only under the supervision of certified clinicians at a health facility[13]. This restriction was suspended temporarily in July 2020 during the public health emergency presented by the COVID-19 pandemic[14–16] and was lifted permanently on December 16, 2021[17]. This opened the door for more people to access medication abortion and other relevant health services via telemedicine, ultimately resulting in increased access to fully remote abortion services via telemedicine in some states, including California[18]. However, some states have now blocked the use of telemedicine for abortion services[19]. According to the Guttmacher Institute, 14 states no longer allow the provision of abortion services. During 2021 and 2022, different states adopted a total of 158 abortion restrictions, 108 in 2021 and 50 in 2022[20], and the attempt for further restrictions has continued to increase during 2023. Despite the adoption of various abortion-related restrictions in recent years and the removal of federal protection for abortion services by the Supreme Court, abortion remains a common practice in many states. In 2020, a total of 930,160 abortions were performed in the United States, which represents an increase of 8% increase from those performed in 2017[21].

This scoping review systematically mapped studies conducted on abortion services through digital healthcare and telemedicine services, including e-counseling, e-consenting, e-prescribing, and evaluated their effectiveness, success rate, acceptability for potential abortion users and the perspectives of both abortion users and providers on abortion services offered via telemedicine.

[1]The University of Tennessee Health Science Center (UTHSC) - Oak Ridge National Laboratory (ORNL) Center for Biomedical Informatics, Department of Pediatrics, College of Medicine, Memphis, TN 38103, USA. ✉email: fkumsa@uthsc.edu; ashabann@uthsc.eduPublished in partnership with Seoul National University Bundang Hospital    npj

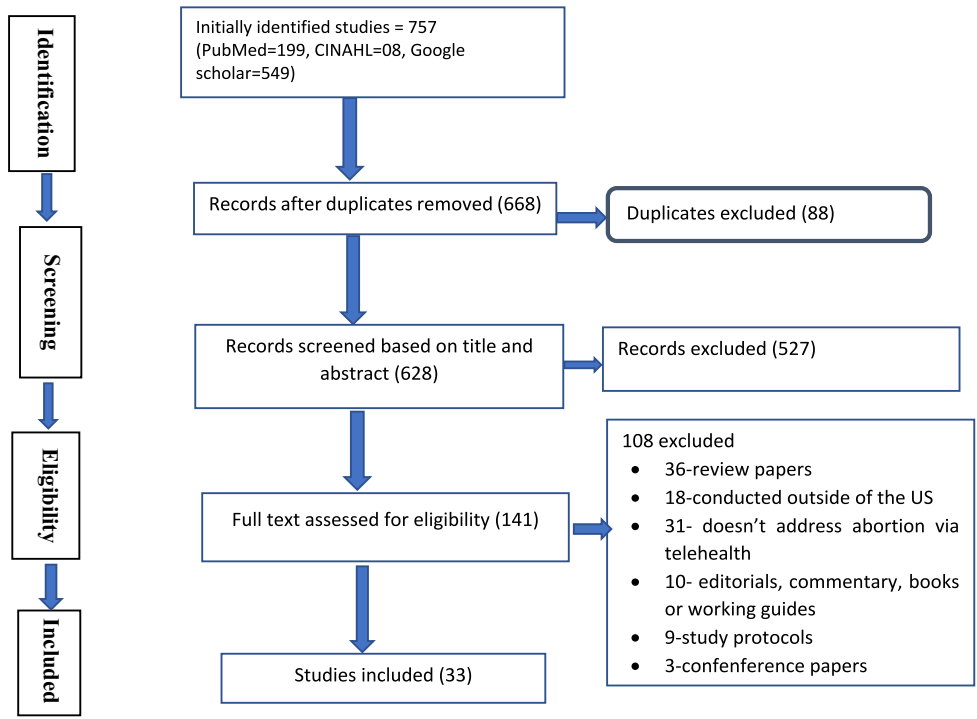

**Fig. 1** Articles screening process regarding digital health and telemedicine access to medication abortion in the United States, 2023.

## RESULTS
Our articles screening strategy for this scoping review is shown in Fig. 1. We identified a total of 757 studies and removed 88 duplicates and 527 other articles based on information in the title and abstract. The remaining 141 studies were subjected to full-text review, and 108 were excluded for reasons that included failure to report medication abortion through telemedicine as an outcome variable, studies contacted outside of the United States, or review papers. The final scoping review included 33 studies (Fig. 1).

The characteristics of the included studies are shown in Table 1. Of the studies included in the final review, 11 were qualitative studies[11,12,15,22–29], 6 were cross-sectional studies[30–35], and 13 were cohort (prospective, retrospective, or follow-up) studies[8,10,16,18,36–44]. One study used mixed methods approach[45], one used a multicenter single-arm clinical trial[46], and one used an investigational new drug application (INDA) approach[13]. The participants in 26 studies were potential abortion users[7,8,10,11,14–16,18,22,27,28,31–34,36–46], the participants of two studies included both abortion users and providers[26,29], and those of the remaining five studies were abortion care providers[12,23–26,35] (Table 1).

### Telemedicine and abortion access
The results and themes identified from the included studies are shown in Table 2. Six studies assessed the role of telemedicine in increasing access to abortion care[25,32,34,38,41,43], particularly for potential abortion users living in remote areas[38]. Grossman et al. (2013) reported that the proportion of potential abortion users who used medication abortion increased from 46% to 54% following the introduction of telemedicine[38]. Similarly, Kohn et al. (2021) showed that the proportion of medication abortion users in Montana increased from 60% to 65%, and the number of abortion users for those in Nevada increased from 461 to 735 after the implementation of abortion via telemedicine[41]. Finally, a study by Thompson et al. (2021) suggested that making abortion care available via telemedicine could increase the abortion access rate from 11.1 to 12.3 per 1000 reproductive-age women[34] (Table 2).

### Providers' view and experience of telemedicine
Six studies explored the views and experiences of abortion care providers regarding the provision of medication abortion via telemedicine[12,23–26,29]. These providers expressed their feelings that telemedicine had expanded the access to medication abortion for their patients[29] and that telemedicine needed to adopt processes similar to those used in in-person clinic visits, with minor additional technological arrangements to facilitate the electronic interface between patients and doctors[12].

Qualitative studies of the providers' experiences suggested that the provision of medication abortion via telemedicine facilitated a more user-centered approach, with abortion seekers receiving the services closer to their homes, thereby reducing the need for long-distance travel by both patients and physicians[12,26] and providing flexible appointment times[26]. However, in a study conducted among telehealth leaders, a few participants expressed concerns that in providing abortion services via telemedicine, care providers would be unable to verify their patient's identity, would experience difficulties in ensuring that the abortion medication provided via telemedicine would be taken by the right patient, and the potential that abortion services might be accessed by minors in the absence of parental consent[24].

### Abortion users' views on abortion via telemedicine
Seven studies assessed the views and perceptions of potential abortion users regarding their access to abortion services via telemedicine[11,15,26,28,29,33,45]. Overall, the perceptions of potential abortion users regarding telemedicine abortion services were positive[26]. Many users who had abortions supported the use of telemedicine for medication abortion, particularly during the COVID-19 pandemic period, and felt that the effectiveness of telemedicine service was almost equivalent to that of in-person abortion service[45]. A survey conducted among 1567 abortion users reported that 56% of overall participants and 64% of

Table 1. Characteristics of studies included in digital health and medicine for medication abortion in the United States, 2023.

| Author, year of publication | Dates of data collection | Study design | Study participants | Sample size | Study objective |
|---|---|---|---|---|---|
| Aiken et al. 2020[36] | October 2017 to August 2018 | Record review | Potential abortion users | 6022 | To assess demand and geographical variation in demand for medication abortion through an online telemedicine service. To examine motivations for seeking this service and how types of motivation for doing so vary by state abortion policy context |
| Aiken et al. 2021[30] | March 2018 to March 2020 | Cross-sectional study | Potential abortion users | 57,506 | To examine reasons for accessing medication abortion via online telemedicine. To examine the associations between state- and county-level factors and the rate of requests for abortion service via telemedicine |
| Anger et al. 2021[46] | March to September 2020 | Multicenter single-arm trial | Potential abortion users | 410 | To compare outcomes among patients who did or did not have a pre-abortion ultrasound or pelvic exam before obtaining medication abortion via direct-to-patient telemedicine and mail. |
| Beardsworth et al. 2022[31] | October 2016 to December 2019 | Cross-sectional study | Potential abortion users | 204 | To compare miles and days until medication abortion via Tele-Abortion versus in-clinic |
| Chong et al. 2021[14] | May 2016 to September 2020 | Investigational New Drug application | Potential abortion users | 1157 | To present updated evidence on the safety, efficacy, and acceptability of a direct-to-patient telemedicine abortion service |
| Daniel et al. 2020[10] | January 2015 to March 2018. | Retrospective cohort study | Potential abortion users | 9175 | To evaluate demographic and service delivery differences between patients using telemedicine relative to an in-person visit |
| Ehrenreich et al. 2019[11] | April to October 2017 | Qualitative study | Potential abortion users | 18 | To describe women's experiences using telemedicine for their abortion information visits, including their perceptions and reactions |
| Ehrenreich et al. 2019[22] | May to October 2017 | Qualitative study | Potential abortion users | 20 | To explores women's experiences of using telemedicine for information visits to fulfill abortion requirements and the acceptability of telemedicine to provide information |
| Fiastro et al. 2022[23] | November to December 2020 | Qualitative study | Clinicians and administrators | 21 | To examines rapid innovation of remote abortion service operations across healthcare settings. To describe the use of telehealth consultations with medications delivered directly to patients |
| Fix et al. 2018[24] | June 2014 to February 2015 | Qualitative study | Leaders of U.S. telemedicine and telehealth organizations | 19 | To explore the knowledge and attitudes of telehealth organization leaders towards telemedicine for medication abortion and bans on this use, and to situate these restrictions in the context of other telemedicine services |
| Godfrey et al. 2021[24] | November to December 2020 | Qualitative study | Healthcare providers and administrators | 21 | To identify organizational factors that promoted the successful implementation of telehealth and adoption of "no test" medication abortion protocols |
| Grindlay et al. 2013[26] | October 2009 to February 2010 | Qualitative study | Healthcare providers and potential abortion users | 40 | To evaluate patients' and providers' experiences with telemedicine provision of medical abortion |
| Grindlay et al. 2017[12] | October to November 2013 | Qualitative study | Healthcare providers | 8 | To evaluate providers' experiences with telemedicine provision of medication abortion in Alaska |

**Table 1 continued**

| Author, year of publication | Dates of data collection | Study design | Study participants | Sample size | Study objective |
|---|---|---|---|---|---|
| Grossman et al. 2011[37] | November 2008 to October 2009 | Prospective cohort study | Potential abortion users | 578 | To estimate the effectiveness and acceptability of telemedicine provision of early medication abortion compared with the provision of a face-to-face physician visit |
| Grossman et al. 2013[38] | July 2006 to June 2008 versus July 2008 to June 2010 | Retrospective record review | Potential abortion users | 17,801 | To assess the effect of a telemedicine model providing medication abortion on service delivery in a clinic system |
| Grossman et al. 2017[7] | July 2008 to June 2015 | Retrospective cohort study | Potential abortion users | 19170 (8765 telemedicine and 10,405 in-person) | To compare the proportion of medical abortions with a clinically significant adverse event among telemedicine and in-person patients |
| Johnson et al. 2021[27] | June to August 2019 | Qualitative study | Potential abortion users | 80 | To examine the impact of economic circumstances on abortion care pathway decision-making and experiences seeking care |
| Kaller et al. 2021[32] | March 2017 to March 2018 | Cross-sectional study | Potential abortion users | 383 (166 telemedicine and 217 in-person) | To identify factors associated with using telemedicine for informed consent, patients' reasons for using it, and experiences with it, compared to in-person informed consent |
| Kapp et al. 2021[39] | 2016 to 2019 | Retrospective records review | Potential abortion users | 144 | To evaluate medication abortion effectiveness and safety in women at 13 or more weeks gestation provided care through Women on Web's telemedicine service |
| Kerestes et al. 2021[15] | January to July 2020 | Qualitative study | Potential abortion users | 45 | To understand how obtaining a medication abortion by mail with telemedicine counseling versus traditional in-clinic care impacted participants' access to care |
| Kerestes et al. 2021[40] | April to November 2020 | Retrospective cohort study | Potential abortion users | 334 | To demonstrate the effectiveness of medication abortion with the implementation of telemedicine and a no-test protocol in response to the COVID-19 pandemic |
| Kohn et al. 2019[42] | April 2017 to March 2018 | Retrospective records review | Potential abortion users | 5952 (738 telemedicine and 5214 standards) | To assess outcomes of medication abortion provided through telemedicine compared with standard medication abortion |
| Kohn et al. 2021[41] | 2015 to 2018 | Retrospective records review | Potential abortion users | 3038 (1314 pre- vs. 1724 postimplementation) | To assess changes in service delivery patterns 1 year after introducing telemedicine for medication abortion |
| LaRoche et al. 2021[45] | October to November 2020 | Mixed methods study | Potential abortion users | 332 | To explore public opinion about using telemedicine to provide medication abortion during the COVID-19 pandemic in 2020. |
| Madera et al. 2022[28] | June to August 2019 | Qualitative study | Potential abortion users | 80 | To examine people's motivations and experiences searching for, self-sourcing, and completing a self-managed medication abortion at home using Aid Access |
| Pleasants et al. 2022[43] | August 2017 to May 2018 | Prospective cohort study | Potential abortion users | 1485 | To investigate the association of distance to the nearest abortion facility with abortion or pregnancy outcome |

| Table 1 continued | | | | | |
|---|---|---|---|---|---|
| Author, year of publication | Dates of data collection | Study design | Study participants | Sample size | Study objective |
| Raymond et al. 2019[16] | May 2016 to December 2018 | Follow-up study | Potential abortion users | 433 | To evaluate the safety, feasibility, and acceptability of a direct-to-patient telemedicine service that enabled people to obtain medication abortion without visiting an abortion provider in person. |
| Rivlin et al. 2022[33] | April 2020 to April 2021 | Cross-sectional study | Potential abortion users | 1218 | To assess associations between mistrust in the abortion provider and preferences for a telemedicine abortion |
| Ruggiero et al. 2022[29] | June 2017 to July 2019 | Qualitative study | Healthcare providers and potential abortion users | 29 | To document the experiences of patients and providers using telehealth for medication abortion |
| Thompson et al. 2021[34] | April 2018 to October 2019 | Cross-sectional study | Potential abortion users | 2015 | To examine the association between travel distance to the nearest abortion care facility and the abortion rate and to model the effect of reduced travel distance. |
| Upadhyay et al. 2020[35] | April to May 2020 | Survey | Abortion providers | 120 | To understand management strategies adopted by clinics in response to the COVID-19 pandemic. |
| Upadhyay et al. 2021[18] | October 2020 to January 2021 | Retrospective cohort study | Potential abortion users | 141 | To assess the safety and efficacy of medication abortions by telehealth in the US during the COVID-19 pandemic |
| Wiebe et al. 2020[44] | January 2017 to January 2019 | Retrospective chart review | Potential abortion users | 381 (182 telemedicine and 199 in-clinic) | To compare the practical aspects of providing medication abortions through telemedicine and in-person clinic visits |

medication abortion participants preferred to receive abortion services via telemedicine[33].

According to the study conducted by Kerestes et al. (2021), potential abortion users reported that the availability of abortion services through telemedicine made these services more accessible and convenient relative to the services provided in-clinic[15]. The study further elaborated that abortion services via telemedicine were the option preferred by patients because their receipt of counseling via telemedicine and abortion medication by mail eliminated a long waiting time for appointments and the need for other arrangements, including transportation and childcare required for in-clinic service[15]. Similarly, abortion users in another study reported that using abortion services via telemedicine helped them to minimize the financial, travel, and time-related burdens necessitated by in-person visits[11]. However, a significant number of potential abortion users expressed worry about the time-sensitive nature of abortion in case the medication should not arrive within the recommended time[28].

The majority of the abortion users reported that addressing their concerns and medical-related questions to the care providers over telemedicine was similar to or more comfortable than face-to-face visits[26,29]. However, some users felt less personally and emotionally connected with the care providers on telemedicine video calls, felt that video calls were less legitimate than in-person visits, or expressed concerns about scams and breaches of privacy and therefore preferred to be in the same room with their abortion care providers[11,26,29,45]. Moreover, several participants of the study described by Ehrenreich et al. (2019) reported that they did not find the information script provided through telemedicine to be informative, either because of its content or delivery method (i.e., felt that using telemedicine to deliver the information would be weird)[11].

### Reason for choosing abortion services via telemedicine

Ten studies reported the reasons that potential abortion users elected to use telemedicine rather than face-to-face meetings for their abortion services[11,12,15,16,22,26,27,30,32,36]. The most common reasons for preferring telemedicine included the long distance from their homes to the abortion clinics[11,16,22,30,36], convenience[11,15,16,32], privacy[11,16,22,30,36], cost (unaffordable cost of in-clinic abortion or personal financial hardship)[11,16,27,30,36], the flexibility of appointment times[16,32,36], state laws that included a waiting period or specified restrictive policies for accessing an abortion clinic[22,27,36], perceived stigma[16,22,36], and preference for talking via a video call[32]. Two studies by Aiken et al. (2020, 2021) reported that 69–73% of potential abortion users who selected the use of telemedicine did so because of their inability to afford in-clinic abortion services, while 39–49% did so for reasons of privacy[30,36]. On the other hand, Kaller et al. (2021) reported that 73% of the users they surveyed chose to receive their abortion services via telemedicine because it was more convenient than a clinic visit[32].

### Effectiveness of abortion via telemedicine

Ten studies investigated the effectiveness of medication abortion services offered via telehealth in the United States[7,14,16,18,27,37,39,40,42,44]. Kerestes et al. (2021) reported that 97% of telemedicine users completed their abortion without requiring additional surgical intervention, compared to 93.6% of clinic visit patients[40]. Similarly, Grossman et al. (2011) reported that 99% of abortions among telemedicine users were successful, while the success rate was 97% for face-to-face patients[37].

The majority of the studies we reviewed reported that ≤6% of abortions were completed without the need for additional surgical intervention[14,16,18,40,42,44]. The rates of clinically adverse effects reported ranged from none to ≤ 1%[7,18,37,42], and no deaths were reported[7,42]. However, a study by Women on Web conducted

Table 2. Results or themes identified from the included studies in medication abortion via telemedicine in the United States, 2023.

| Author, year of publication | Results or themes |
|---|---|
| Aiken et al. 2020[36] | ✓ 76% lived in hostile, and 24% lived in supportive states<br>✓ Reason for requesting telemedicine abortion (74% expressed more than one reason)<br>  ○ 69.1% due to cost, 39.2% due to privacy, 34.0% due to time flexibility, 27.1% due to distance to a clinic, 17.1% due to state law (e.g., waiting period), and 16.4% due to perceived stigma.<br>✓ Preference for abortion through telemedicine<br>  ○ 49.1% for privacy, 46.4% for comfort when used at home, 42.3% for autonomy feeling, 25.4% for the ability to have others present, and 11.3% for a feeling of empowerment.<br>✓ Barriers to using telehealth for potential users living in states supportive of abortion compared to the hostile states.<br>  ○ Cost (63% vs. 71%), distance to abortion provider facility (21% vs. 29%), abortion-related legal restrictions (14% vs. 18%), and protestors (12% vs. 15%) |
| Aiken et al. 2021[30] | ✓ Reasons for requests abortion via telehealth<br>  ○ Not affording in-clinic abortion service (73.5%)<br>  ○ Privacy (49.3%)<br>  ○ Distance from the abortion-providing clinic (40.4%)<br>✓ County-level factors associated with requests.<br>  ○ Distance to the closest abortion-providing clinic<br>  ○ The proportion living below the federal poverty level |
| Anger et al. 2021[46] | ✓ 18 participants needed additional intervention to complete the abortion or for the pregnancy to be continued. Of these,<br>  ○ 5.6% no-test vs. 1.6% test-medication abortion<br>  ○ 4.2% no-test vs. 0 % test group had procedural interventions<br>  ○ No statistically significant differences in adverse outcomes or ongoing pregnancy after medication abortion |
| Beardsworth et al. 2022[31] | ✓ Time to mifepristone administration or ingestion was 11 days for tele abortion while 6 days for in-clinic patients ($p < 0.01$)<br>✓ The median distance to the abortion clinic was 7 miles for each group |
| Chong et al. 2021[14] | ✓ 95% of the abortion were completed without a procedure<br>✓ 6% of the participants made unplanned emergency visits for urgent care<br>✓ 0.9% of the participants developed serious adverse events such as hospitalization or blood transfusions<br>✓ Enrollment in telemedicine abortion care substantially increased following the COVID-19 occurrence<br>✓ 99% were satisfied with the service they received |
| Daniel et al. 2020[10] | ✓ 91% received abortion services in-clinic, while 9% received them via telemedicine<br>✓ Compared to in-clinic abortion users, telemedicine abortion users were more likely<br>  ○ Older (27 vs. 25 years)<br>  ○ Live out of state (47% vs. 4%)<br>  ○ Live far away from the abortion clinic (104 vs. 10 miles).<br>No significant differences regarding days taken to receive abortion service following the informed consent visit between the groups |
| Ehrenreich et al. 2019[11] | Three themes were identified:<br>✓ Reasons for choosing telemedicine: Convenience<br>✓ Experience using telemedicine: feeling relieved, worried, preferred an in-person appointment, emotionally unable to connect with care providers<br>✓ Reactions to Information Visit and Waiting Period Requirements: felt impersonal, difficulties due to waiting for time law, and emotional and logistical burdens |
| Ehrenreich et al. 2019[22] | Three themes:<br>✓ Temporal dimensions of abortion access: the mandatory 72-h waiting period (to avoid restriction or potential abortion users pursued the service from other states)<br>✓ Material dimension: the lack of nearby facilities (opted for telemedicine largely because of travel distance)<br>✓ Social dimensions: privacy, stereotypes, or negative views |
| Fiastro et al. 2022[23] | ✓ Most facilities did not provide online abortion services before the COVID-19 pandemic.<br>✓ Most clinics included in the study-initiated telemedicine medication abortion services and integrated them into existing in-clinic services |
| Fix et al. 2018[24] | Four themes:<br>✓ Opinions of appropriate uses for telemedicine: services that do not require physical presence or contact are appropriate for telemedicine.<br>✓ Knowledge and opinions of medication abortion provision via telemedicine felt telemedicine well-suited for medication abortion.<br>✓ Knowledge and opinions of telemedicine abortion restrictions:<br>  ○ Had limited knowledge of abortion restriction laws.<br>  ○ Opposed bans on abortion services through telemedicine.<br>✓ Restrictions on other telemedicine services: the broader field of telemedicine experienced similar restrictions due to the requirements for provider-patient-relationships, and difficulties in monitoring the user-controlled substances by remote prescription |
| Godfrey et al. 2021[25] | Implementation of telemedicine abortion services needs<br>✓ Access to inter-organizational networks<br>  ○ Appropriate professional organizations<br>  ○ Mentorship from telemedicine service innovators<br>✓ Readiness of organizations for telemedicine service implementation<br>  ○ Working electronic health records<br>  ○ Options for virtual provider-patient interactions<br>Effective and motivated clinic workers |
| Grindlay et al. 2013[26] | Advantages of telemedicine compared with an in-person provision<br>✓ It decreases physicians' and patients' travel<br>✓ Greater availability of different locations and appointment times<br>✓ Abortion users were either indifferent or had positive views about conversation through telemedicine<br>  ○ Some felt private, or secure<br>  ○ Some are even more comfortable than in-person communications<br>Some preferred to be in the same room with care providers |

| Table 2 continued | |
|---|---|
| Author, year of publication | Results or themes |
| Grindlay et al. 2017[12] | Participants experience the provision of medication abortion through telemedicine<br>✓ It facilitated a user-centered service provision approach<br>✓ Integrating new technology into the existing clinical service is easy<br>✓ Impacts of telemedicine on patients:<br>  ○ Flexibility in appointment times<br>  ○ Reduction in patient travel<br>✓ Impacts of Telemedicine on Clinics and Providers<br>  ○ Low impact on patient-provider interaction and clinic flow<br>  ○ Improve the efficiency of the clinic<br>✓ Suggestions for service improvement:<br>  ○ Make the patient not see herself in the monitor<br>  ○ Ensure everyone knows who is in the room<br>Navigate between the medical chart and the patient video |
| Grossman et al. 2011[37] | ✓ The proportion of successful abortions was 99% (96–100%) for telemedicine and 97% (94–99%) for in-person patients<br>✓ 91% were very satisfied with their abortion service<br>✓ 25% of telemedicine patients preferred to be in the same room with care providers.<br>✓ the prevalence of adverse events was similar among telemedicine and in-person abortion service users (1.3%) |
| Grossman et al. 2013[38] | ✓ The proportion of medication abortion users increased from 46% to 54% after the introduction of telemedicine<br>✓ Abortion users traveled a 3.16-mile lower average distance to access abortion services after the telemedicine introduction<br>✓ The proportion of abortion users who lived > 25 miles from a surgical abortion clinic increased from 40% to 44%, while those who lived >50 miles from a surgical abortion clinic increased from 24% to 26% |
| Grossman et al. 2017[7] | ✓49 adverse outcomes were reported<br>  ○ 0.18% (0.11–0.29%) among telemedicine<br>  ○ 0.32% (0.23–0.45%) among in-person patients<br>  ○ No deaths or surgery was reported |
| Johnson et al. 2021[27] | Reasons for telemedicine abortion use was:<br>  ○ The unaffordable cost of in-clinic abortion<br>  ○ Personal financial hardship<br>  ○ Restrictive policies to access the clinic |
| Kaller et al. 2021[32] | ✓ Telemedicine patients would have traveled further distance than an in-person patient to receive abortion services (65 vs. 21 mean miles)<br>✓ The odds of being very satisfied with the visit (aOR, 2,89; 95% CI: 1.93–4.32) and "very comfortable" asking questions during the visit (aOR, 3.76; 95% CI: 2.58–5.49) were higher among telemedicine patients compared to the in-person one.<br>✓ Reasons for choosing telemedicine over the in-person visit<br>  ○ Convenience (73%)<br>  ○ Shorter appointment time (16%)<br>  ○ Preference to talk over video (11%) |
| Kapp et al. 2021[39] | ✓ 29% reported adverse events such as heavy bleeding or fever<br>✓ 43% received additional care from the care providers<br>✓ 18% completed through dilatation and curettage<br>✓ 10% of pregnancies continued |
| Kerestes et al. 2021[15] | Identified themes were:<br>  ○ Telemedicine was more accessible and convenient<br>  ○ Privacy concerns were ameliorated by telemedicine<br>  ○ Participants perceived a lack of alternatives to telemedicine abortion<br>  ○ Telemedicine abortion is highly acceptable<br>  ○ During the COVID-19 pandemic, the advantages of telemedicine abortion were magnified |
| Kerestes et al. 2021[40] | ✓ 44.6% received abortion services via telemedicine, but they picked up abortion medication in person<br>✓ 22.5% received abortion services via telemedicine and abortion pill sent by mail<br>✓ 32.9% received abortion services via traditional in-person visits<br>✓ 95.8% of abortions completed without additional surgical intervention Success rates<br>  ○ 96.8% for telemedicine users but picked up from the clinic<br>  ○ 97.1% for telemedicine users but received by mail<br>  ○ 93.6% for a traditional in-clinic visit<br>  ○ 96.6% without an ultrasound performed before the abortion<br>  ○ 95.5% for with ultrasound before the abortion procedure |
| Kohn et al. 2019[42] | Among patients with follow-up data<br>  ○ 0.5% of the pregnancy continued for telemedicine users, while 1.8% of pregnancies continued for in-clinic users.<br>  ○ 1.4% of the abortion completes by aspiration procedures for telemedicine users, while 4.5% are completed by aspiration for in-clinic users<br>  ○ adverse events reported from <1% of each group<br>  ○ No deaths reported |
| Kohn et al. 2021[41] | After the implementation of telemedicine<br>In Montana<br>  ○ The proportion of abortion users who utilized medication abortion services increased from 60% to 65%<br>  ○ The mean appointment time for abortion services decreased from 14 to 12 days<br>  ○ The mean distance traveled (one way) for abortion service decreased from 134 to 115 miles<br>  In Nevada<br>  ○ The number of abortion users who utilized medication abortion services increased from 461 to 735<br>  ○ The mean distance traveled (one way) for abortion services decreased from 47 to 34 miles |

| Table 2 continued | |
|---|---|
| Author, year of publication | Results or themes |
| LaRoche et al. 2021[45] | ○ 44% of the participants supported the use of telemedicine for medication abortion during the pandemic<br>○ 35 % opposed<br>○ 21% were not sure about the decision<br>Themes identified:<br>○ Safety-related perceptions were tied to attitudes toward using telemedicine for medication abortion<br>○ Participants had a concern about the legitimacy of telemedicine for medication abortion<br>○ Participants felt that abortion should be carried out as early as possible |
| Madera et al. 2022[28] | Themes identified:<br>○ Viewed telemedicine as a "godsend"<br>○ Shipping delays, fears of scams, and surveillance made ordering pills online a "nerve-racking" experience<br>○ A "personal touch" calmed fears and fostered trust in telemedicine<br>○ Worried about the "what ifs" of abortion through telemedicine experience<br>○ Online telemedicine met their needs |
| Pleasants et al. 2022[43] | Living ≥50 miles away from the facility providing abortion care was associated with seeking an abortion or planning to continue pregnancy four weeks later |
| Raymond et al. 2019[16] | Three hundred sixty-three were scheduled for telemedicine abortion evaluations with clinicians<br>○ 89% reported convenience, cost, speed, or lack of options at their locality as reasons for choosing telemedicine service<br>○ 13% specifically reported cost as a reason for choosing telemedicine<br>○ 21% reported privacy concerns as a reason<br>• 94% of abortions are completed by surgical or medical interventions.<br>• Four abortion users had ongoing pregnancies after ingestion of mifepristone<br>• Five abortion users developed bleeding |
| Rivlin et al. 2022[33] | ✓ 546 used medication abortion services<br>✓ 56% of all participants reported that they prefer telemedicine service<br>✓ 64% of abortion users who received medication abortion preferred telemedicine services<br>✓ 1.4% of all participants mistrusted abortion service providers, while 1% of medication abortion users mistrusted service provider |
| Ruggiero et al. 2022[29] | ✓ The majority of the telemedicine users felt comfortable talking to care providers over telemedicine and faced no problems while using the telemedicine technology<br>✓ Health care providers believe that telemedicine expanded access to medication abortion service |
| Thompson et al. 2021[34] | ✓ The analysis included 3107 counties, which had a total of 62.5 million reproductive-age females<br>✓ The estimated mean abortion rate per 1000 reproductive-age female residents was 11.1 [1.0–45.5]<br>✓ Integrating abortion care into primary health care would increase the mean abortion rate to 11.4 [1.1–45.5] per 1000 reproductive-age females, resulting in an additional 18,190 abortions<br>✓ Widely availing telemedicine services for abortion care would further increase the mean abortion rate to 12.3 [1.4–45.5] per 1000 reproductive-age females, resulting in an additional 70,920 abortions |
| Upadhyay et al. 2020[35] | ✓ 87% changed abortion-related protocols due to COVID-19<br>○ 71% moved follow-ups to telemedicine modalities such as video or phone<br>○ 41% initiated or increased telemedicine service for screening and consultations<br>○ 43% reduced Rh testing<br>○ 42% reduced other tests<br>○ 15% omitted the pre-abortion ultrasound requirements<br>✓ 20% of the facilities allowed quick pickup of abortion pills<br>✓ 4% of the facilities began mailing abortion pills to patients after a telemedicine consultation<br>✓ Clinical practice changes such as starting or increasing telemedicine services were reported across the U.S.<br>✓ The clinical practice changes were higher in the Northeast (73%) compared to clinics located in the South (23%) |
| Upadhyay et al. 2021[18] | ✓ 95% of the abortion completed without surgical intervention<br>✓ No patients developed any major adverse events |
| Wiebe et al. 2020[44] | ✓ 18.1% of telemedicine and 100% of in-clinic patients had dating ultrasounds<br>✓ Aspiration was used to complete 3.3% of abortions for telemedicine users and 4.5% for in-clinic patients<br>✓ Lost to follow-up was reported for 5.5% of telemedicine users and 6.6% of in-clinic patients<br>✓ complications were experienced by 5.5% of telemedicine users compared to 5.0% of in-clinic patients<br>✓ Unscheduled communications with office assistants were made by 46.2% of telemedicine users and 21.6% of in-clinic patients |

between 2016 and 2019 among 131 recipients of medication abortion at ≥13 gestational weeks via telemedicine stated that 29% of the abortion users reported adverse events such as fever or heavy bleeding, that 43% received additional care, and that 18% of the abortion procedures occurred by aspiration[39]. Similarly, Raymond et al. (2019) reported that 7% of the participants visited an emergency center for urgent care[16].

Regarding the development of adverse events, almost all studies reported no significant difference in the development of clinically significant adverse effects among telemedicine users and standard or face-to-face abortion care users[37]. However, the use of an aspiration procedure was less common among telemedicine users as compared to standard procedure user patients (1.4% vs. 4.5%)[42]. More patients also used the telemedicine service for unscheduled communications with office assistants than did patients who received their care in-person (46.2% vs. 21.6%)[44].

**Telemedicine abortion service during the COVID-19 pandemic**
A few studies reported how the COVID-19 pandemic influenced the utilization of abortion services via telemedicine[8,14,23,35]. The demand for abortion services by telemedicine greatly increased during the COVID-19 pandemic[8,14,23]. Most clinics integrated telemedicine services into clinical care to supplement existing abortion services in the clinic and for other patient appointments[23].

A study conducted among 100 clinics reported that 87% of the clinics made changes to their service protocols due to the COVID-19 pandemic, including initiating or increasing the use of telemedicine for patient screening, consultations, or follow-up, eliminating or reducing the requirement for pre-abortion testing such as ultrasound and blood tests to screen for Rh factor, and providing rapid access to abortion pills. Facilities in the North reported a higher increase in the use of telemedicine for abortion services (73%) than did facilities in the South (23%)[35].

## DISCUSSION
This scoping review synthesized evidence from 33 studies that describe access to abortion via telemedicine, including its effectiveness, the reasons that abortion users preferred to receive abortion service via telemedicine, and the views and perceptions of abortion users and providers in the United States. The use of telemedicine in general became prevalent during COVID-19, and telemedicine became user-friendly for the provision of medication abortion services[8,14,23]. The present review determined that telemedicine increased access to abortion care in the country, especially for potential abortion users who lived in remote places[16,23,31]. Care providers felt that telemedicine increased access to medication abortion, facilitated a more user-centered approach, and reduced long-distance traveling time by both patients and physicians. Although telemedicine is believed to make abortion services more accessible, the study conducted in Brazil revealed inequalities in the utilization of abortion services via telemedicine across various regions, which varied based on income, and among different racial groups[47]. Given that provision of services via telemedicine requires access to advanced technologies such as high internet speed, populations who lack reliable internet access or who lack knowledge and education about the use of these technologies could be disproportionally affected by limited ability to utilize the service. The identified gaps indicate that further studies will be needed to explore the feasibility of using abortion services through telemedicine, particularly in low-resource settings and among racialized groups.

The present scoping review points out that abortion services via telemedicine are highly acceptable by both potential abortion users and care providers, a finding that was also reported by another review[48]. Participants and care providers believed that the provision of abortion service via telemedicine facilitated a pregnant person-centered approach, provided flexibility in appointment times, and reduced costs and travel requirements by patients and physicians. Pregnant persons showed several layers of conflicting emotions about receiving abortion services via the modality of telemedicine. Most abortion users preferred to receive abortion services via telemedicine due to its convenience, low cost, and their belief that the telemedicine service could maintain confidentiality. However, a significant number of abortion seekers raised concerns that talking over a video made them feel impersonal, that they felt difficulty in connecting emotionally with the care providers over telemedicine, and that the information script they received was not informative. Waiting for mifepristone to arrive via mail also created a delay in abortion initiation[31]. Although telemedicine is believed to have increased access to abortion services and is considered convenient by many, especially during the COVID-19 pandemic, it is important not to overlook the patients' concerns lest they be deterred from effectively utilizing telemedicine services. Further research could also be needed on how to address these concerns.

The effectiveness of telemedicine in providing abortion services was measured in various ways, including complete uterine evacuation without additional medical or surgical intervention, the absence of adverse effects such as heavy bleeding, blood transfusion, hospitalization, and death, and no longer being pregnant. Several studies reported the success rate of abortion service via telemedicine, with ≤6% of the abortions being completed with the help of additional surgical intervention[14,16,18,40,42,44], and only a few patients experiencing severe adverse outcomes. Abortion service via telemedicine was reported to be as effective as abortion services provided in-clinic[49,50]. This is consistent with the results of a study conducted outside of the United States[51].

This study possesses some limitations. Firstly, the limited number of available studies and the inconsistent reporting across these studies prevented us from effectively synthesizing and pooling some of the findings. In addition, we were unable to discern the influence of socio-economic and race-related factors on the utilization of telemedicine for abortion. In conclusion, this scoping review aims to provide a comprehensive synthesis of findings from published literature on abortion services via telemedicine in the United States. The findings of this review showed that telemedicine increased access to abortion care in the United States, particularly for those who resided in remote areas or who feared the potential stigma of receiving abortion care from the clinic. Studies reported that there was no significant difference regarding the effectiveness of abortion offered via telemedicine and in-person clinic visits. Overall, both abortion care providers and users had positive perceptions of abortion services via telemedicine service. However, findings from the potential abortion users show that they felt a range of conflicting emotions about abortion services via the modality of telemedicine. The majority of the abortion users preferred telemedicine over in-person visit abortion services, while some users preferred to be physically present in the same room with their care providers. As digital technologies, including telemedicine, drive the health industry toward more comprehensive equitable solutions[52] and interactive tailored platforms for care navigation and delivery, there are still notable gaps in the availability, access, quality, and affordability of these technologies in delivering care related to abortion, miscarriage, and other related issues. To maintain or increases the acceptability of abortion service via telemedicine, addressing the participants' concerns about the use of telemedicine should not be overlooked.

## METHODS
### Protocol
This study employed a scoping review protocol consistent with the guidelines of the Preferred Reporting Items for Systematic Reviews and Meta-Analyses Extension for Scoping Reviews (PRISMA-ScR)[53]. After the review was conducted, the four phases of the PRISMA flow chart were completed to show the screening process we used.

### Article Search
We used Google Scholar to conduct an exhaustive search for primary studies in the United States that were published in the English language. Additionally, MeSH terms and CINHAL headings were used to identify references in the PubMed and CINHAL databases. The search strategy combined two main concepts: medication abortion and telemedicine. Similar keywords and vocabulary were combined using the Boolean terms OR and AND between the two concepts. The search terms emerged from "medication abortion," "self-managed abortion," "misoprostol," "tele abortion," "telemedicine," "telehealth," and "United States." The article search was conducted in September 2022.

### Eligibility criteria
All primary studies that addressed the issue of medication abortion via telemedicine in the United States were included, including those using qualitative, quantitative, and mixed-method approaches. Medication abortion via telemedicine includes e-counseling, e-consenting, and/or e-prescribing abortion pills for users residing in the United States, regardless of the service provider's location. Participants for the included studies were either potential abortion users or care providers. The studies we included also examined the effectiveness, acceptability, and perspectives of providers and users of abortion services offered via telemedicine. We excluded ineligible studies based on our exclusion criteria, including duplicate studies, conference papers, anonymous reports, and review papers, and did not report any finding related to medication abortion via telemedicine, editorials, and theoretical papers.

### Screening
First, we searched for relevant articles using specific search terms and applied filters. To facilitate the screening process, we exported the selected articles to Covidence, a web-based systematic review software (Veritas Health Innovation, Melbourne, Australia; available at www.covidence.org). After removing the duplicates, the studies were screened based on the inclusion criteria using the information included in titles and abstracts. Finally, we conducted a full-text review of the eligible papers to determine their inclusion. Two authors (FK and RP) handled the article screening process. Any disagreement between these two authors on the inclusion or exclusion of the articles was resolved through discussion with ASN, who oversaw the screening process and the overall activities of this review.

### Data charting process and synthesis
We followed the recommendations for scoping reviews in performing data extraction, analysis, and presentation of results, using a Microsoft Excel spreadsheet for data extraction[54]. The list of authors, study objective, design, participants, data collection time, publication year, sample size, and outcome of interest was abstracted, and the major study characteristics and their findings were summarized using a table. The findings were synthesized narratively. Data extraction and synthesis of the finding was performed by F.K., while A.S.N. and R.P. reviewed the synthesized draft and provided their critical comments.

### Reporting summary
Further information on research design is available in the Nature Research Reporting Summary linked to this article.

## DATA AVAILABILITY
The authors declare that the data supporting the findings of this study are available within the paper.

## ACKNOWLEDGEMENTS
We would like to thank the University of Tennessee Health Science Center (UTHSC)—Oak Ridge National Laboratory (ORNL) Center for Biomedical Informatics for providing the resources needed for conducting this research. This research received no specific grant from any funding agency in the public, commercial, or not-for-profit sectors.

## AUTHOR CONTRIBUTIONS
F.K. conceived the study, conducted the literature search, screened the papers, and drafted the manuscript. R.P. screened the studies and critically reviewed the manuscript. A.S.N. has also conceived the study, oversaw the overall process, critically reviewed and edited the manuscript, supervised the study, and acquired the funding. All authors read this paper and finally approved it for submission.

## COMPETING INTERESTS
The authors declare no competing interests.

## ADDITIONAL INFORMATION
**Supplementary information** The online version contains supplementary material available at https://doi.org/10.1038/s41746-023-00871-2.

**Correspondence** and requests for materials should be addressed to Fekede Asefa Kumsa or Arash Shaban-Nejad.

**Reprints and permission information** is available at http://www.nature.com/reprints

**Publisher's note** Springer Nature remains neutral with regard to jurisdictional claims in published maps and institutional affiliations.